# A geometric scaling between collective organizations and interaction-space dimension

Arturo Tozzi (corresponding author)
ASL Napoli 1 Centro, Distretto 27, Naples, Italy
Via Comunale del Principe 13/a 80145
tozziarturo@libero.it

## ABSTRACT

The number of stable macroscopic organizations in complex systems is often much smaller than the large number of microscopic degrees of freedom would suggest. Yet theoretical approaches rarely address whether general limits constrain the diversity of admissible macroscopic organizations. We develop a geometric framework in which interactions among system components define a coarse-grained interaction space endowed with a metric structure. When this space has finite intrinsic dimensionality, geometric packing constraints impose bounds on the number of mutually distinguishable collective organizations. We derive a dimension-dependent scaling law showing that the number of stable macroscopic regimes grows polynomially with exponent equal to the intrinsic dimensionality of the interaction space. This implies that increasing microscopic complexity alone does not necessarily expand the range of macroscopic organizations. Instead, diversification requires an increase in the dimensionality of effective interactions. To illustrate our approach, we analyze an interacting system in which collective regimes correspond to regions of a low-dimensional parameter space describing effective interactions. In this setting, geometric packing constrains the number of robust organizations that the system can support. Overall, we argue that dimensionality of interaction space may act as a control parameter governing a variety of collective organization across physical and biological systems.

## 1. INTRODUCTION

Collective organization emerges when interacting components produce coherent macroscopic structures through local interactions. This behavior appears across numerous nonlinear systems, yet the diversity of stable macroscopic organizations is often surprisingly limited. High-dimensional microscopic descriptions frequently collapse into a small number of dominant collective variables capturing the relevant macroscopic behavior (Becht et al. 2018; Thrun and Ultsch 2020; Okuma et al. 2021; Malepathirana et al. 2022; Jeon et al. 2025). Systems with different microscopic structures converge toward common large-scale descriptions governed by a small set of relevant parameters (Kadanoff 1966; Wilson 1971; Liao and Polonyi 1995; De Polsi et al. 2020; Cagnetta et al. 2022; Dolai, Simha and Basu 2024). Related ideas appear in studies of self-organized criticality and catastrophe theory, where nonlinear systems evolve toward marginal states characterized by reduced parameter spaces and scale-invariant dynamics (Thom 1975; Bak et al. 1987; Bogdan and Wales 2004; Merli and Pavese 2018; Gabriele et al. 2020; Ather 2021). In complex systems theory, hierarchical organization and near-decomposability restrict viable interactions between subsystems, thereby limiting feasible collective regimes (Simon 1962; Esteve-Altava et al. 2015; Kronfeldner 2021; Rivelli 2025). Related geometric ideas also appear in quantum many-body physics, where spatial structure may emerge from patterns of entanglement (Swingle 2012; Sokolov et al. 2022; Joshi et al. 2023). In theoretical biology, approaches emphasizing self-organization, morphogenetic fields and systemic causation suggest that only a subset of conceivable biological organizations is physically realizable (Kauffman 1993; Goodwin 2001; Noble 2006; Kerstjens et al., 2025). These observations motivate the question addressed in this work: whether the geometry and dimensionality of interaction spaces impose general quantitative limits on the diversity of collective organizations sustained by nonlinear systems.

We propose a geometric framework in which interactions among system components are represented as points in a coarse-grained interaction space endowed with a metric structure. If interactions were compressed into a bounded region of a space with finite intrinsic dimensionality, geometric packing constraints could impose limits on the number of mutually distinguishable collective organizations. We hypothesize that the diversity of robust organizations might follow a dimension-dependent scaling law determined by the intrinsic dimensionality of the interaction space. Diversification might require an expansion of the dimensionality of effective interactions, corresponding to the introduction of additional independent interaction modes.
We analyze an interacting system in which effective interactions are described by a small number of parameters controlling interaction strength, spatial range and directional coupling. Distinct macroscopic organizations correspond to separated regions of a low-dimensional interaction space and their number is limited by geometric packing constraints. More generally, we shift attention from specific generative mechanisms to geometric constraints on the structure of interaction spaces.

## 2. INTERACTION SPACES WITH COARSE-GRAINED GEOMETRY AND GEOMETRIC SEPARABILITY

We introduce an interaction space to describe the effective interactions among system components after coarse graining. Coarse-graining procedures replace microscopic variables with effective descriptors, summarizing the net influence that components exert on one another. These descriptors may correspond to reduced coupling parameters, interaction kernels or aggregated measures derived from microscopic dynamics. The set of all admissible descriptors defines the interaction space in which collective organizations are represented.

Let $\mathcal{J}$ denote the space of effective interaction descriptors for a given class of systems. Each point $x \in \mathcal{J}$ represents a complete specification of interactions at the chosen level of coarse graining. Different microscopic systems may correspond to the same point in $\mathcal{J}$ whenever their macroscopic interaction structures are equivalent. To compare interaction structures quantitatively, we assume that $\mathcal{J}$ is endowed with a metric $d(\cdot,\cdot)$ that measures the distinguishability between two interaction descriptors. The distance $d(x, y)$ expresses how strongly the effective interactions represented by $x$ differ from those represented by $y$. These metrics arise naturally from normed parameter spaces and operator norms on interaction kernels induced by coarse-graining procedures. The specific metric must satisfy the standard properties of positivity, symmetry and triangle inequality.

We assume that the interaction space has finite intrinsic dimensionality. Formally, we assume that $\mathcal{J}$ admits an embedding into a metric manifold of intrinsic dimension $d$. This dimension characterizes the number of independent directions along which effective interactions can vary. Crucially, $d$ needs not coincide with the number of microscopic degrees of freedom in the underlying system, which may be arbitrarily large. Instead, it reflects the dimensionality of the coarse-grained interaction geometry governing macroscopic organization. The assumption of low intrinsic dimensionality is supported by empirical observations. Coarse-grained descriptions in physics, biology and data analysis frequently reveal that interactions collapse into a small set of dominant parameters or modes, even when microscopic complexity is enormous. In these cases, the relevant region of interaction space occupies a bounded subset of a low-dimensional manifold.

We therefore consider bounded subsets $\Omega \subset \mathcal{J}$ representing the range of interaction descriptors accessible to a given class of systems. Boundedness reflects the physical requirement that effective interaction strengths remain finite after coarse graining. Within these regions, the geometry of $\mathcal{J}$ determines how many distinct interaction configurations can be separated by a finite distance. Therefore, our geometric formulation allows collective organization to be analyzed independently of detailed dynamics.

When the interaction space has finite intrinsic dimension, purely geometric considerations impose universal constraints on the number and separability of distinct collective organizations. Once defined the interaction space $\mathcal{J}$, we now formalize the notion of collective organization in a way that does not depend on the specific microscopic dynamics. The aim is to identify the minimal geometric structure required to express limits on the diversity of macroscopic organizations.

We define a collective organization as a macroscopic regime stable under small perturbations of effective interactions. In the interaction-space representation, each organization corresponds to a subset $\mathcal{O} \subset \mathcal{J}$ containing all interaction descriptors generating indistinguishable macroscopic behavior at the level of coarse-grained observables. These observables may include spatial patterns, dynamical regimes or stable structural configurations, depending on the system under consideration. Different microscopic realizations may correspond to the same organization if their effective interactions lie within the same subset of $\mathcal{J}$.

To characterize distinct organizations geometrically, we introduce the notion of separability in interaction space. Two organizations $\mathcal{O}_1$ and $\mathcal{O}_2$ are said to be geometrically separable if there exists a positive scale $\varepsilon > 0$ such that
$$d(x, y) \geq \varepsilon \text{ for all } x \in \mathcal{O}_1, y \in \mathcal{O}_2.$$

This condition expresses the requirement that the two organizations remain distinguishable under perturbations of interactions smaller than $\varepsilon$. Therefore, geometric separability captures the notion of robust macroscopic difference: organizations that are separated in this sense cannot be transformed into one another by arbitrarily small changes in effective interactions.

A family of organizations $\{\mathcal{O}_1, \mathcal{O}_2, \ldots, \mathcal{O}_N\}$ is said to coexist within an interaction region $\Omega \subset \mathcal{J}$ if each organization corresponds to a subset of $\Omega$ and all pairs are geometrically separable. This coexistence does not require simultaneous realization in physical space, rather expresses the ability of the interaction framework to support multiple distinct macroscopic regimes without mutual interference in parameter space.

Therefore, the problem of organizational diversity becomes a geometric question. Each robust organization corresponds to a region of interaction space and distinct organizations must occupy regions that are separated by a finite distance. The maximal number of these organizations depends on how many separated regions can be accommodated within the bounded interaction domain $\Omega$. When the intrinsic dimensionality of $\mathcal{J}$ is finite, this becomes a classical geometric packing problem, in which the number of mutually separable regions that can be placed within a bounded metric space is limited by its dimensionality and metric properties. As a result, the diversity of collective organizations cannot grow arbitrarily within a fixed interaction geometry.

In the following section, we derive a dimension-dependent scaling law to bound the number of robust organizations coexisting in a finite-dimensional interaction space.

## 3. DIMENSIONAL SCALING OF ORGANIZATIONAL DIVERSITY

We now derive our central result: when effective interactions occupy a bounded region of a finite-dimensional interaction space, the diversity of mutually distinguishable collective organizations is constrained by a geometric scaling law determined by the intrinsic dimensionality of that space.

Let $\mathcal{J}$ denote the interaction space introduced above and let $\Omega \subset \mathcal{J}$ be the bounded region of interaction descriptors accessible to a given class of systems. Assume that $\Omega$ has diameter $R$ with respect to the metric $d(\cdot,\cdot)$.

Consider a family of collective organizations that are pairwise geometrically separable with separation scale $\varepsilon > 0$. Each organization contains at least one representative interaction descriptor such that the distance between representatives of distinct organizations satisfies

$$d(\theta_i, \theta_j) \geq \varepsilon \quad (i \neq j).$$

Under this condition, the set of representative interaction descriptors forms an $\varepsilon$-separated set in $\Omega$.

The problem of determining the maximal number of these organizations can be formulated as a geometric packing problem: how many points separated by at least $\varepsilon$ can be placed within a bounded region of a space with intrinsic dimension $d$. This leads to the following result.

**Theorem (Dimension–diversity scaling law).**
Let $(\Omega, d)$ be a bounded interaction space of intrinsic dimension $d$ and diameter $R$. Suppose that collective organizations correspond to regions of interaction space whose representative descriptors form an $\varepsilon$-separated set in $\Omega$. Then, the maximal number $N$ of mutually distinguishable organizations satisfies

$$N \leq C_d \left(\frac{R}{\varepsilon}\right)^d,$$

where $C_d$ is a constant depending on the geometric properties of the interaction space.

**Proof.**
Because the interaction space has intrinsic dimension $d$, any bounded subset admits a finite $\varepsilon$-packing number. The maximal number of points that can be placed in a region of diameter $R$ while maintaining pairwise distance at least $\varepsilon$ scales as

$$N_{\max} \sim \left(\frac{R}{\varepsilon}\right)^d.$$

Since each robust collective organization must occupy a region of interaction space separated from others by distance $\varepsilon$, the same packing bound applies to the maximal number of organizations. ∎

This result follows from standard packing arguments in finite-dimensional metric spaces and establishes a quantitative relationship between the diversity of collective organizations and the intrinsic dimensionality of interaction space. For fixed robustness scale $\varepsilon$, the maximal number of distinguishable organizations grows polynomially with exponent $d$.
Consequently, increasing microscopic complexity does not automatically produce a greater variety of macroscopic organizations. Even systems containing extremely large numbers of microscopic degrees of freedom remain constrained if their effective interactions collapse into a low-dimensional manifold. Still, the emergence of new collective organizations requires an expansion of interaction dimensionality. Introducing additional independent interaction parameters or modes increases the dimensionality of the interaction space, enlarging the region available for geometric packing.

These observations lead to a dimension–diversity principle: the maximal diversity of robust collective organizations supported by a system scales with the intrinsic dimensionality of its interaction space. The repertoire of admissible macroscopic organizations is controlled not by the microscopic system's size, but by the geometry of coarse-grained interactions.
In the next section we examine how this scaling relation behaves when interaction spaces evolve in time or when interactions are asymmetric or nonreciprocal.

The derived scaling law assumes that the interaction space is static and that the metric describing distinguishability between interaction descriptors remains fixed. However, in many real systems effective interactions evolve in time or display asymmetries that violate reciprocity. Examples include adaptive biological networks, active matter systems with energy injection and many-body systems whose interaction structure changes dynamically. It is therefore useful to determine whether the dimensional scaling constraint remains valid under these more general conditions.
We first consider the case of time-dependent interaction spaces. Let $\mathcal{J}(t)$ denote a family of interaction spaces parameterized by time, each equipped with a metric $d_t(\cdot,\cdot)$. We assume that the intrinsic dimensionality of the interaction space remains finite and constant over time and that the metric evolves continuously, so that interaction descriptors do not undergo arbitrarily large discontinuous displacements in infinitesimal time intervals. Under these conditions, the bounded region of interaction space accessible to the system deforms smoothly as interactions evolve. If collective organizations are defined through regions of interaction space separated by a scale $\varepsilon$, continuous deformation of the interaction geometry cannot generate an unbounded number of new organizations. The reason is geometric: continuous deformation preserves the finiteness of packing numbers in bounded regions of finite-dimensional spaces. Consequently, the maximal number of mutually separable organizations is bounded by a relation of the same form as the scaling law derived above.

A second extension concerns nonreciprocal or asymmetric interactions. In many biological and active systems, the influence of component $i$ on component $j$ is not necessarily equal to the influence of $j$ on $i$. These asymmetries can arise from directed signaling, energy-consuming processes or active transport mechanisms. At the level of effective descriptors, these interactions may naturally be represented by asymmetric kernels or matrices. Even in these cases the geometry of interaction space can still be characterized by a distance measure capturing distinguishability between interaction structures. A convenient approach is to define a symmetrized distance

$$d(x,y) = \frac{1}{2}(d_{\text{dir}}(x,y) + d_{\text{dir}}(y,x)),$$

where $d_{\text{dir}}$ represents an asymmetric distance associated with direct interactions. The resulting metric preserves the intrinsic dimensionality of the interaction space, allowing geometric separation between organizations to be defined in the same manner as before. Because the dimensionality of the symmetrized metric space remains finite, the packing argument leading to the dimension–diversity scaling law still applies.

Therefore, the constraint linking organizational diversity to interaction-space dimensionality is robust under a broad class of extensions. Neither continuous temporal evolution of interactions nor the presence of nonreciprocal couplings is sufficient to circumvent the geometric limits imposed by low-dimensional interaction spaces. Only processes that effectively increase the dimensionality of interactions by introducing new independent modes or interaction channels, can expand the number of distinct collective organizations that a system can support.

In the following section, we examine the implications of this geometric scaling principle across different classes of physical and biological systems.

## 4. IMPLICATIONS ACROSS CLASSES OF SYSTEMS: AN ILLUSTRATIVE EXAMPLE

The dimensional scaling relation derived above has implications for systems in which collective behavior emerges from many interacting components. Because our argument relies only on the geometric structure of coarse-grained interaction spaces, the resulting constraints hold independently of the detailed microscopic dynamics governing the system.

In passive many-body systems like interacting particle assemblies, polymers or granular materials, coarse graining often compresses complex microscopic interactions into a small number of effective parameters. Under these conditions, our dimensional scaling law implies that the diversity of robust macroscopic organizations is limited by the dimensionality of the interaction space rather than by the number of microscopic components. Related structural restrictions have been reported in granular systems, where coarse-grained variables evolve toward steady states occupying restricted regions of configuration space despite different microscopic dynamics (Wanjura et al. 2019). For active and nonequilibrium systems, including self-propelled agents or driven particle assemblies, energy injection can generate complex dynamics; however, if the effective interactions remain confined to a low-dimensional manifold, the same geometric constraints persist. Continuous temporal deformation of interaction geometry or nonreciprocal couplings does not remove the packing limits imposed by finite interaction dimensionality. Consequently, transitions between collective regimes may correspond to changes in the geometry or dimensionality of interaction space rather than increases in microscopic complexity alone.

Similar considerations arise in molecular biological systems. Studies of heteropolymer and protein interactions show that effective interaction landscapes between disordered protein sequences collapse into low-dimensional parameter spaces, limiting the number of sequences capable of robust demixing (Adachi and Kawaguchi 2024). Although biological networks involve many molecular components, their regulatory interactions are often governed by a small set of collective variables. Our dimensional scaling law predicts that the number of stable organizational regimes, such as differentiated states or functional compartments, cannot grow indefinitely within a fixed interaction architecture. Related robustness phenomena have been observed in hierarchical biological materials, where additional structural levels increase tolerance to assembly errors without expanding the diversity of configurations (Michel and Yunker 2019). Likewise, limits on chromatin condensates have been attributed to mechanical frustration and geometric restrictions imposed by the chromatin environment (Zhang et al. 2021).

These observations suggest that the restricted repertoire of stable phase organizations may arise from geometric constraints on effective interaction dimensionality rather than detailed biochemical properties.

To illustrate how the dimensional scaling principle operates in practice, we analyze a simple nonlinear system in which collective organization emerges from pairwise effective interactions among agents. Consider a population of $N$ interacting agents with positions $x_i(t)$. System evolution is governed by the dynamical equation

$$\frac{dx_i}{dt} = \sum_{j \neq i} J_{ij} (x_j - x_i),$$

where $J_{ij}$ denotes the effective interaction kernel between agents $i$ and $j$. We assume that the kernel takes the form

$$J_{ij} = \alpha \exp\left(-\frac{|x_i - x_j|}{\beta}\right) + \gamma A_{ij},$$

where $\alpha$ represents the interaction strength, $\beta$ determines the spatial interaction range and $\gamma$ controls an alignment or directional coupling term represented by $A_{ij}$. Similar parameterizations arise frequently in coarse-grained descriptions of interacting particle systems and collective motion models.

At the coarse-grained level, the interaction structure of the system is specified by the parameter vector

$$\theta = (\alpha, \beta, \gamma).$$

The admissible interaction descriptors form a three-dimensional interaction space

$$\mathcal{J} \subset \mathbb{R}^3.$$

Assume that the parameters vary within bounded intervals

$$\alpha \in [0, \alpha_{\max}], \beta \in [0, \beta_{\max}], \gamma \in [0, \gamma_{\max}],$$

so that the accessible interaction domain occupies a bounded region $\Omega \subset \mathcal{J}$ with characteristic diameter $R$.

Depending on the values of $\theta$, the nonlinear system may exhibit homogeneous mixing, spatial clustering, oscillatory collective motion or phase-separated domains. Each regime corresponds to a region of interaction space whose points produce similar macroscopic behavior under the system dynamics.

To quantify the distinction between interaction descriptors, we introduce the Euclidean metric

$$d(\theta_1, \theta_2) = \|\theta_1 - \theta_2\|.$$

Two collective organizations are considered distinct when their corresponding regions in interaction space are separated by a distance larger than a robustness scale $\varepsilon$, such that the representative descriptors associated with distinct organizations form an $\varepsilon$-separated set within the bounded domain $\Omega$.

Because the interaction space has intrinsic dimension $d = 3$, the maximal number of mutually separable organizations that can coexist within $\Omega$ is constrained by the geometric packing bound

$$N_{\max} \leq C_3 \left(\frac{R}{\varepsilon}\right)^3,$$

where $C_3$ is a constant determined by the metric properties of the interaction space. The resulting dependence of organizational diversity on the dimensionality of interaction space is illustrated in Figure.

This example illustrates the mechanism proposed in this work. Even when the underlying system contains many microscopic degrees of freedom, the diversity of robust collective organizations remains limited because effective interactions vary only within a low-dimensional parameter space. Therefore, increasing the number of possible organizations requires expanding the dimensionality of the interaction descriptors, for example by introducing additional independent interaction parameters or interaction channels. When effective interactions are governed by a small number of coarse grained parameters, the number of stable macroscopic regimes that the system can support remains intrinsically limited, despite potentially huge microscopic complexity.

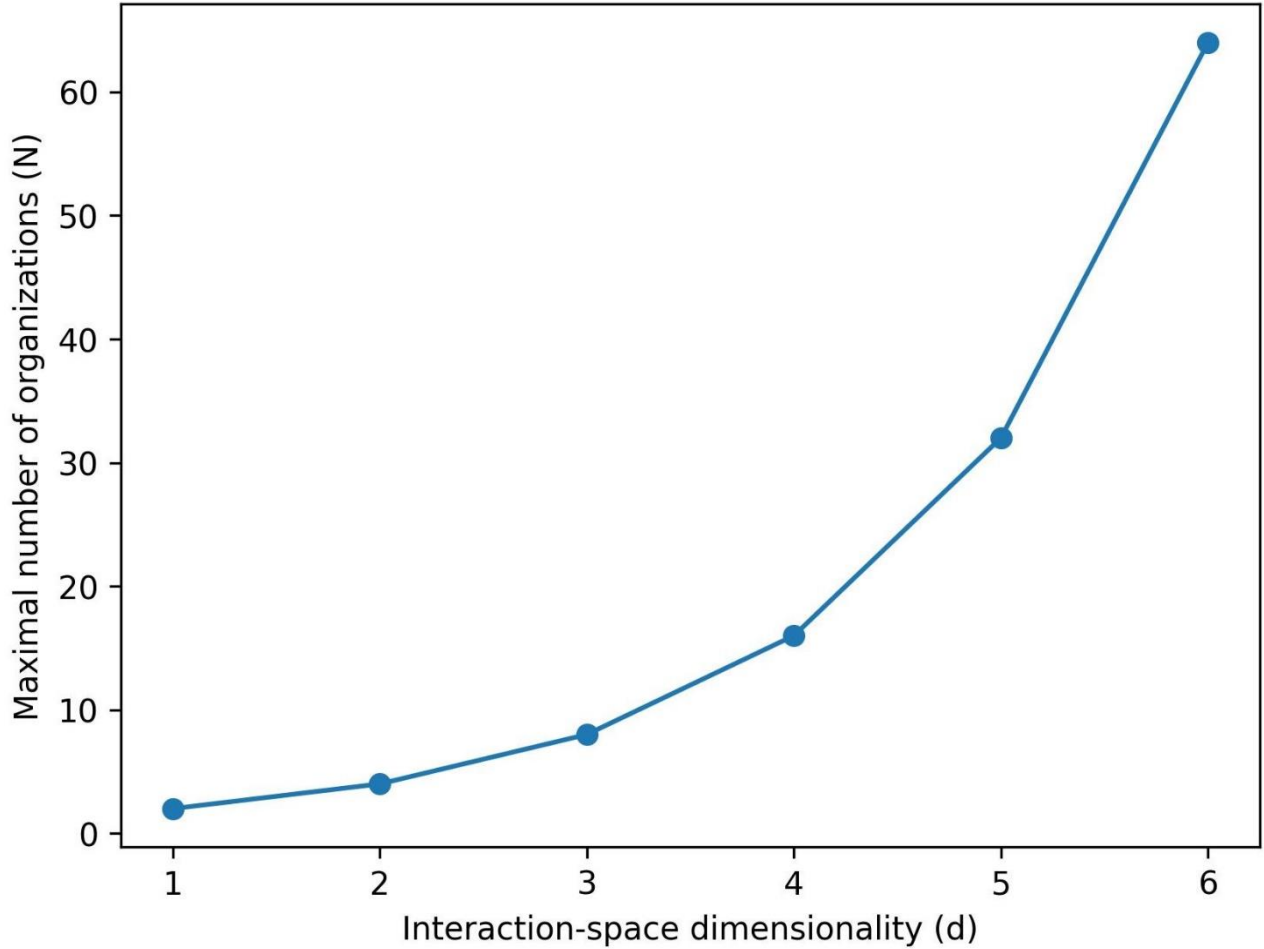

**Figure.** Dimensional scaling of organizational diversity. The maximal number of distinguishable and stable collective organizations increases polynomially with the intrinsic dimensionality of the coarse-grained interaction space. Higher-dimensional interaction spaces can accommodate more separated regions corresponding to distinct organizations, consistent with the scaling relation $N \leq C_d (R/\varepsilon)^d$. When interactions are confined to a low-dimensional space, only a limited number of robust collective regimes can exist.

## 5. CONCLUSIONS

We developed a geometric framework for understanding limits on collective organization in complex systems. By representing effective interactions as points in a coarse-grained interaction space endowed with a metric structure, we showed that the diversity of robust macroscopic organizations is constrained by the intrinsic dimensionality of that space. When effective interactions occupy a bounded region of a finite-dimensional manifold, the number of mutually distinguishable collective organizations obeys a dimension-dependent scaling law derived from geometric packing arguments.

Our approach establishes a direct relationship between organizational diversity and interaction-space dimensionality. For a fixed robustness scale separating distinct organizations, the maximal number of stable macroscopic regimes grows polynomially with exponent equal to the intrinsic dimension of the interaction space. This implies that, if effective interactions are confined to a low-dimensional manifold, increasing the number of microscopic components or the complexity of microscopic dynamics does not necessarily increase the diversity of macroscopic organizations. Distinct macroscopic organizations correspond to separated regions of a low-dimensional interaction space, their number being limited by geometric packing constraints.

We followed a well-established practice in physics and complex systems research: using known mathematical results to derive new physical interpretations and constraints on system behavior. Our contribution does not lie in the inequality itself, but in its interpretation within the context of collective organization. Rather than a new theorem of metric geometry, our approach provides a conceptual bridge between the geometry of interaction space, the diversity of macroscopic organizations and the range of collective regimes that nonlinear systems can support.

Traditional approaches to collective organization focus on identifying the dynamical mechanisms that generate peculiar macroscopic patterns. In contrast, we address the complementary question of which organizations are fundamentally impossible given the geometry of effective interactions. From this perspective, the intrinsic dimensionality of interaction space acts as a control parameter governing the repertoire of admissible collective structures.

Several directions for future work follow naturally. On the theoretical side, sharper bounds may be obtained by incorporating additional geometric structure into the interaction space, like curvature or measure-theoretic properties. On the empirical side, estimating the effective dimensionality of interaction spaces in concrete systems could provide a route for testing the predicted scaling relation between dimensionality and organizational diversity.

Our approach suggests that many limits on collective organization arise not from the details of microscopic dynamics, but from geometric constraints inherent in the structure of effective interactions. Identifying and quantifying these constraints may contribute to explain why the diversity of macroscopic organization observed across complex systems is far smaller than the space of conceivable possibilities.

**DECLARATIONS**

**Ethics approval and consent to participate.** This research does not contain any studies with human participants or animals performed by the Author.
**Consent for publication.** The Author transfers all copyright ownership, in the event the work is published. The undersigned author warrants that the article is original, does not infringe on any copyright or other proprietary right of any third part, is not under consideration by another journal and has not been previously published.
**Availability of data and materials.** All data and materials generated or analyzed during this study are included in the manuscript. The Author had full access to all the data in the study and took responsibility for the integrity of the data and the accuracy of the data analysis.
**Competing interests.** The Author does not have any known or potential conflict of interest including any financial, personal or other relationships with other people or organizations within three years of beginning the submitted work that could inappropriately influence or be perceived to influence their work.
**Funding.** This research did not receive any specific grant from funding agencies in the public, commercial or not-for-profit sectors.
**Acknowledgements:** none.
**Authors' contributions.** The Author performed: study concept and design, acquisition of data, analysis and interpretation of data, drafting of the manuscript, critical revision of the manuscript for important intellectual content, statistical analysis, obtained funding, administrative, technical and material support, study supervision.
**Declaration of generative AI and AI-assisted technologies in the writing process.** During the preparation of this work, the author used ChatGPT 4o to assist with data analysis and manuscript drafting and to improve spelling, grammar and general editing. After using this tool, the author reviewed and edited the content as needed, taking full responsibility for the content of the publication.